\documentstyle[12pt]{article}
\begin{document}
\hfill{NCKU-HEP-98-13}\par
\hfill{hep-ph/9810222}\par
\vfill
\centerline{\large{\bf Perturbative QCD analysis of exclusive}}\par
\centerline{\large{\bf $B$ meson decays\footnote{talk presented
at the Fourth Workshop on Particle Physics Phenomenology, Kaohsiung,
Taiwan, 1998}}}\par
\vskip 1.5cm
\centerline{Hsiang-nan Li }
\vskip 0.5cm
\centerline{Department of Physics, National Cheng-Kung University,}
\centerline{Tainan, Taiwan 701, Republic of China}
\vskip 1.0cm
%PACS numbers: 13.20He, 12.15Hh, 12.38Bx
\vskip 2.0 cm
%\baselineskip=2\baselineskip

\centerline{\bf Abstract}
\vskip 0.5cm

We review the perturbative QCD formalism for exclusive heavy meson decays,
concentrating on the three-scale factorization theorem for nonleptonic
processes. The formalism is then extended to the radiative decay
$B\to K^*\gamma$, which occurs through penguin diagrams. It is observed
that the contributions from the operators other than the penguin one
$b\to s\gamma$ are not negligible. From the best fit to the experimental
data of the branching ratio ${\cal B}(B\to K^*\gamma)$, we extract the $B$
meson wave function, which possesses a sharp peak in the region with a
small momentum fraction.

\vfill

\newpage
\centerline{\large \bf 1. Introduction}
\vskip 0.5cm

Recently, perturbative QCD (PQCD) has been proposed to be an alternative
approach to the study of heavy hadron decays \cite{LY1,L1,LY3}, which
complements the heavy quark effective theory \cite{G} and the
Bauer-Stech-Wirbel (BSW) method \cite{BSW}. In this talk we review the PQCD
analysis of exclusive $B$ meson decays, concentrating on the three-scale
factorization theorem for nonleptonic processes \cite{CL}. The PQCD
formalism is then applied to the radiative decay $B\to K^*\gamma$, which
involves the contributions from penguin diagrams. According to the
factorization theorem, a heavy meson decay rate is expressed as the
convolution of a hard subamplitude with meson wave functions. The former is
calculable in usual perturbation theory, while the latter must be extracted
from experimental data or derived by using nonperturbative methods, such as
QCD sum rules. Since the $K^*$ meson wave function has been known from the
sum rule analyses \cite{CZ}, the $B\to K^*\gamma$ decay is an ideal
process, from which the unknown $B$ meson wave function can be determined.
With the $B$ meson wave function obtained here, we are able to make
predictions for other $B$ meson decay modes.

The operator for exclusive semileptonic heavy meson decays, such as
$B\to D^{(*)} l\nu$, is
\begin{equation}
\frac{G_F}{\sqrt{2}}V_{cb}{\bar \nu}\gamma_{\mu}(1-\gamma_5)l
{\bar c}\gamma^{\mu}b\;,
\end{equation}
where $G_F$ is the Fermi coupling constant and $V_{cb}$ the
Cabibbo-Kabayashi-Maskawa (CKM) matrix element. The formalism for
semileptonic decays is simpler, and part of that for nonleptonic decays.
Hence, we shall explain only the factorization of nonleptonic heavy meson
decays below, which occur through the Hamiltonian,
\begin{eqnarray}
H=\frac{G_F}{\sqrt 2}V_{ij}V_{kl}^{*}({\bar q}_l q_k)
({\bar q}_j q_i)\;,
\label{full}
\end{eqnarray}
$({\bar q} q)={\bar q} \gamma_\mu(1-\gamma_5)q$ being the $V-A$ current.
Hard gluon corrections cause an operator mixing, and their
renormalization-group summation leads to the effective Hamiltonian,
\begin{eqnarray}
H_{\rm eff}=\frac{G_F}{\sqrt 2}V_{ij}V_{kl}^{*}[c_1(\mu)O_1(\mu)+ 
c_2(\mu)O_2(\mu)]\;,
\label{eff}
\end{eqnarray}
with the four-fermion operators $O_1=({\bar q}_l q_k)({\bar q}_j q_i)$ and 
$O_2=({\bar q}_j q_k)({\bar q}_l q_i)$. The Wilson coefficients $c_{1,2}$, 
organizing the large logarithms from the hard gluon corrections to all 
orders, describe the evolution from the $W$ boson mass $M_W$ to a lower
scale $\mu$ with the matching conditions $c_1(M_W)=1$ and $c_2(M_W)=0$.

The most widely adopted approach to exclusive nonleptonic heavy meson
decays is the BSW model \cite{BSW}, in which the factorization hypothesis
on the matrix elements of the operators $O_{1,2}$ is assumed. In this model
decay rates are expressed in terms of various hadronic transition form
factors. Employing the Fierz transformation, the coefficient of the form
factors corresponding to external $W$ boson emissions is $a_1=c_1+c_2/N$,
and that corresponding to internal $W$ boson emissions is $a_2=c_2+c_1/N$,
$N$ being the number of colors. The form factors may be related to each
other by heavy quark symmetry, and parametrized by different ansatz.
Nonfactorizable contributions, which can not be expressed in terms of
hadronic transition form factors, and nonspectator contributions from
$W$ boson exchanges are neglected.

Though the BSW model is simple and gives predictions in fair agreement with 
experimental data, it encounters several difficulties. It has been known 
that the large $N$ limit of $a_{1,2}$, {\it i.e.}, the choice 
$a_{1}=c_{1}(M_c)\approx 1.26$ and $a_{2}=c_{2}(M_c)\approx -0.52$,
with $M_c$ the $c$ quark mass, explains the data of charm decays \cite{BSW}. 
However, the same large $N$ limit of $a_{1}=c_{1}(M_b)\approx 1.12$ and 
$a_{2}=c_{2}(M_b)\approx -0.26$, $M_b$ being the $b$ quark mass, does not 
apply to the bottom case. That is, the different mechanism between
charm and bottom decays can not be understood in the BSW approach. Moreover,
it has been difficult to explain the two ratios associated with the
$B\to J/\psi K^{(*)}$ decays \cite{GKP},
\begin{equation}
R=\frac{{\cal B}(B\to J/\psi K^*)}{{\cal B}(B\to J/\psi K)}\;,\;\;\;\;
R_L=\frac{{\cal B}(B\to J/\psi K_L^*)}{{\cal B}(B\to J/\psi K^*)}\;,
\label{rrl}
\end{equation}
simultaneously using the BSW model, where ${\cal B}(B\to J/\psi K)$ is
the branching ratio of the decay $B\to J/\psi K$. This controversy was
found to be attributed to the neglect of the nonfactorizable
internal $W$-emission amplitudes, which are in fact of the same order as
the factorizable ones \cite{CM,WYL2}.

The above difficulties have been overcome by a modified PQCD formalism, the
three-scale factorization theorem \cite{CL}. In the next section we shall
summarize the basic ideas of this theorem.

\vskip 1.0cm

\centerline{\large \bf 2. Three-scale factorization theorem}
\vskip 0.5cm

Nonleptonic heavy meson decays involve three scales: the $W$ boson mass
$M_W$, at which the matching conditions of the effective Hamiltonian to the
original Hamiltonian are defined, the typical scale $t$ of a hard
subamplitude, which reflects the dynamics of heavy meson decays, and the
factorization scale $1/b$, with $b$ the conjugate variable of parton
transverse momenta. The dynamics below $1/b$ is regarded as being
completely nonperturbative, and parametrized into a meson wave fucntion
$\phi(x)$, $x$ being the momentum fraction. PQCD for the scale above $1/b$
is reliable, and radiative corrections produce two types of large
logarithms $\ln(M_W/t)$ and $\ln(tb)$. The former are summed to give the
evolution from $M_W$ down to $t$ described by the Wilson coefficients
$c(t)$. While the latter are summed to give the evolution from $t$ to $1/b$.

There exist also double logarithms $\ln^2(Pb)$ from the overlap of
collinear and soft divergences, $P$ being the dominant light-cone component
of a meson momentum. The resummation of these double logarithms leads to a
Sudakov form factor $\exp[-s(P,b)]$, which suppresses the long-distance
contributions in the large $b$ region, and improves the applicability of
PQCD around the energy scale of few GeV. The $b$ quark mass scale is
located in the range of applicability. This is the motivation we develop
the PQCD formalism for heavy hadron decays. For the detailed derivation of
the Sudakov factor for heavy meson decays, refer to \cite{LY1,L1}. With all
the large logarithms organized, the remaining finite contributions are
absorbed into a hard $b$ quark decay subamplitude $H(t)$. Because of
Sudakov suppression, the perturbative expansion of $H$ in the coupling
constant $\alpha_s$ makes sense.

Therefore, a three-scale factorization formula possesses the typical
expression, 
\begin{eqnarray}
c(t)\otimes H(t)\otimes \phi(x)
\otimes\exp\left[-s(P,b)-2\int_{1/b}^t\frac{d{\bar\mu}}
{\bar\mu}\gamma_q(\alpha_s({\bar\mu}))\right],
\label{for}
\end{eqnarray}
where the exponential containing the quark anomalous dimension $\gamma_q$
describes the evolution from $t$ to $1/b$ mentioned above. As the hard
scale $t$ runs to below $M_b$ and $M_c$, the constructive and destructive
interferences involved in bottom and charm decays, respectively, appear
naturally. Furthermore, not only factorizable, but nonfactorizable
and nonspectator contributions can be evaluated in a systematic way. With
the inclusion of nonfactorizable contributions, we find that $a_{1,2}$
restore their original role of the Wilson coefficients, instead of being
treated as the BSW free parameters. The branching ratios of the various
decay modes $B\to D^{(*)}\pi(\rho)$ and the ratios $R$ and $R_L$ associated
with the $B\to J/\psi K^{(*)}$ decays can all be well explained by the
above formalism \cite{YL}.

We review the evaluation of the factorizable, nonfactorizable and
nonspectator contributions to exclusive nonleptonic decays
$B\to D^{(*)}\pi{(\rho)}$ in the three-scale PQCD factorization theorem.
The factorizable external $W$-emission amplitudes define the $B\to D^{(*)}$
transition form factors $\xi_i$, $i=+$, $-$, $V$, $A_1$, $A_2$, and $A_3$,
through the hadronic matrix elements,
\begin{eqnarray}
\langle D (P_2)|V^\mu|B(P_1)\rangle
&=&\sqrt{M_BM_D}[\xi_+(\eta)(v_1+v_2)^\mu+
\xi_-(\eta)(v_1-v_2)^\mu]\;,
\nonumber \\
\langle D^* (P_2)|V^\mu|B(P_1)\rangle
&=&i\sqrt{M_BM_{D^*}}\xi_V(\eta)
\epsilon^{\mu\nu\alpha\beta}\epsilon^*_\nu v_{2\alpha}v_{1\beta}\;,
\nonumber \\
\langle D^* (P_2)|A^\mu|B(P_1)\rangle&=&\sqrt{M_BM_{D^*}}
[\xi_{A_1}(\eta)(\eta+1)\epsilon^{*\mu}
-\xi_{A_2}(\eta)\epsilon^*\cdot v_1 v_1^\mu
\nonumber \\
& &-\xi_{A_3}(\eta)\epsilon^*\cdot v_1 v_2^\mu]\;.
\label{iwm}
\end{eqnarray}
$P_1$ $(P_2)$, $M_B$ ($M_{D^{(*)}}$) and $v_1$ $(v_2)$ are the momentum, 
the mass, and the velocity of the $B$ $(D^{(*)})$ meson, satisfying the 
relation $P_1=M_Bv_1$ $(P_2=M_{D^{(*)}}v_2)$. $\epsilon^*$ is the 
polarization vector of the $D^*$ meson. The velocity transfer $v_1\cdot v_2$ 
in two-body nonleptonic decays takes the maximal value $\eta=(1+r^2)/(2r)$ 
with $r=M_{D^{(*)}}/M_B$. In the infinite mass limit of $M_B$ and
$M_{D^{(*)}}$ the form factors $\xi_i$ obey the relations
\begin{equation}
\xi_+=\xi_V=\xi_{A_1}=\xi_{A_3}=\xi,\;\;\;\;  \xi_-=\xi_{A_2}=0,
\label{iwr}
\end{equation}
where $\xi$ is the so-called Isgur-Wise (IW) function \cite{IW}.

Equation (\ref{iwm}) is the standard definition of the form factors
$\xi_i$. In our approach, however, the factorization formulas for $\xi_i$ 
will contain the Wilson coefficients as the extra convoluiton factors.
$\xi_i$ include the contributions from the hadronic matrix element of $O_1$ 
and from the color-suppressed matrix element of $O_2$. Therefore, their
factorization formulas involve the Wilson coefficient $a_1=c_1+c_2/N$. 
The form factors for the internal $W$-emission diagrams include the
factorizable contributions from the matrix elements of $O_2$ and from the
color-suppressed matrix element of $O_1$, which then contain the Wilson
coefficient $a_2=c_2+c_1/N$. The form factors for the
$W$-exchange diagrams include the factorizable contributions from the
matrix elements of $O_2$ and from the color-suppressed matrix element of
$O_1$. Hence, they also contain the Wilson coefficient $a_2$. 

The nonfactorizable contributions to the $B\to D^{(*)}\pi$ decays are
color-suppressed, which can not be expressed in terms of hadronic form
factors. A nonfactorizable diagram associated with external $W$
emissions involves a hard exchanged gluon, which, for example, attaches the
spectator quark of the $B$ meson and a valence quark of the pion. The
resultant amplitude depends on the Wilson coefficient $c_2/N$. A
nonfactorizable diagram associated with internal $W$ emissions contains
a hard gluon, which attaches the spectator quark of the $B$ meson and a
valence quark of the $D$ meson. The resultant amplitude depends on
the Wilson coefficient $c_1/N$. Similarly, a nonfactorizable amplitude
associated with $W$ exchanges contains the Wilson coefficient $c_1/N$.
For the explicit expressions of the above form factors and amplitudes,
refer to \cite{YL}.

\vskip 1.0cm

\centerline{\large \bf 3. The $B\to K^*\gamma$ decay}
\vskip 0.5cm

We now apply the three-scale factorization theorem to the radiative
decay $B\to K^*\gamma$ \cite{LL}, whose effective Hamiltonian is 
\begin{equation}
H_{\rm eff}=-\frac{G_F}{\sqrt{2}}V^*_{ts}V_{tb}\sum_{i=1}^{12}
c_i(\mu)O_i(\mu)\;.
\end{equation}
It will be found that only the operators
\begin{eqnarray}
& &O_2=({\bar s}c)({\bar c}b)\;,
\\
& &O_7=\frac{e}{8\pi^2}M_b{\bar s}\sigma^{\mu\nu}(1+\gamma_5)bF_{\mu\nu}\;,
\\
& &O_8=-\frac{g}{8\pi^2}M_b{\bar s}\sigma^{\mu\nu}(1+\gamma_5)T^ab
G^a_{\mu\nu}\;,
\end{eqnarray}
with $\sigma^{\mu\nu}=(i/2)[\gamma^\mu,\gamma^\nu]$, are important. The
$\mu$ dependence of the corresponding Wilson coefficients $c_2(\mu)$,
$c_7(\mu)$ and $c_8(\mu)$ can be quoted from \cite{Buras}.

We assign $P_1=(M_B/\sqrt{2})(1,1,{\bf 0}_T)$ and
$P_2=(M_B/\sqrt{2})(1,r^2,{\bf 0}_T)$ with $r=M_{K^*}/M_B$, $M_{K^*}$
being the $K^*$ meson mass, as the $B$ and $K^*$ meson momenta,
respectively, in the rest frame of the $B$ meson. Assume that the light
valence quark in the $B$ ($K^*$) meson carries the momentum $k_1$ ($k_2$).
$k_1$ has a minus component $k_1^-$, giving the momentum fraction
$x_B=k_1^-/P_1^-$, and small transverse components ${\bf k}_{BT}$. $k_2$
has a large plus component $k_2^+$, giving $x_K=k_2^+/P_2^+$, and
small ${\bf k}_{KT}$. The photon momentum is then $P_3=P_1-P_2$, whose
nonvanishing component is only $P_3^-$.

The decay amplitude is expressed as
\begin{equation}
M=\epsilon_\gamma\cdot\epsilon^*_{K^*}M_1+
i\epsilon_{\mu\rho+-}\epsilon_\gamma^\mu\epsilon^{*\rho}_{K^*}M_2\;,
\label{am}
\end{equation}
with $\epsilon_\gamma$ and $\epsilon_{K^*}$ the polarization vectors
of the photon and of the $K^*$ meson, respectively, and
\begin{equation}
M_{1(2)}=M_{1(2)2}+M_{1(2)7}+M_{1(2)8}\;,
\end{equation}
where the terms on the right-hand side come from the operators $O_2$, $O_7$,
and $O_8$, respectively. From Eq.~(\ref{am}), it is obvious that only the
$K^*$ mesons with transverse polarizations are produced in the decay.
The factorization formulas for the amplitudes $M_{ij}$ are, in terms of 
the overall factor,
\begin{equation}
\Gamma^{(0)}=\frac{G_F}{\sqrt{2}}\frac{e}{\pi}V^*_{ts}V_{tb}C_FM_B^5\;,
\end{equation}
listed below. The $B\to K^*\gamma$ decay rate is then given by
\begin{equation}
\Gamma=\frac{1-r^2}{32\pi M_B}(|M_1|^2+|M_2|^2)\;.
\end{equation}

The amplitude from $O_2$ is written as
\begin{eqnarray}
M_{12}&=&\Gamma^{(0)}\frac{4}{3}\int_0^1dx\int_0^{1-x}dy\int_0^1dx_Bdx_K
\int_0^{1/\Lambda}bdb\phi_B(x_B)\phi_{K^*}(x_K)
\nonumber\\
& &\times \alpha_s(t_2)c_2(t_2)\exp[-S(x_B,x_K,t_2,b,b)]
\nonumber\\
& &\times[(1-r^2+2rx_K+2x_B)y-(rx_K+3x_B)(1-x)]
\nonumber\\
& &\times
\frac{(1-r)(1-r^2)x_Kx}{xy(1-r^2)x_KM_B^2-M_c^2}
H_{2}(Ab,\sqrt{|B_2^2|}b)\;,
\end{eqnarray}
with
\begin{eqnarray}
A^2&=&x_Kx_BM_B^2\;,
\nonumber\\
B_2^2&=&x_Kx_BM_B^2-\frac{y}{1-x}(1-r^2)x_KM_B^2+\frac{M_c^2}{x(1-x)}\;,
\nonumber\\
t_2&=&\max(A,\sqrt{|B_2^2|},1/b)\;,
\end{eqnarray}
where the integrations over $x$ and $y$ are associated with the charm loop,
from which the photon and the hard exchanged gluon are radiated. The
variable $b_B$ ($b_K$), conjugate to the parton transverse momentum
$k_{BT}$ ($k_{KT}$), represents the transverse extent of the $B$ ($K^*$)
meson. $t_2$ is the characteristic scale of the hard subamplitude,
\begin{eqnarray}
H_{2}(Ab,\sqrt{|B_2^2|}b)&=&K_0(Ab)-K_0(\sqrt{|B_2^2|}b)\;,\;\;\;\;
B_2^2>0\;,
\nonumber\\
&=&K_0(Ab)-i\frac{\pi}{2}H_0^{(1)}(\sqrt{|B_2^2|}b)\;,\;\;\;\;
B_2^2<0\;,
\end{eqnarray}
as mentioned before.

There are two diagrams associated with the operator $O_7$, where the hard
gluon attaches the two quarks in the $B$ meson and in the $K^*$ meson
separately. The corresponding amplitude is 
\begin{eqnarray}
M_{17}&=&\Gamma^{(0)}2\int_0^1dx_Bdx_K\int_0^{1/\Lambda}
b_Bdb_Bb_Kdb_K\phi_B(x_B)\phi_{K^*}(x_K)(1-r^2)
\nonumber\\
& &\times\left[rH^{(a)}_{7}(Ab_K,\sqrt{|B_7^2|}b_B,\sqrt{|B_7^2|}b_K)
F_7(t_{7a})\right.
\nonumber\\
& &\left.+[1+r+(1-2r)x_K]H^{(b)}_{7}(Ab_B,C_7b_B,C_7b_K)F_7(t_{7b})\right]\;,
\end{eqnarray}
with
\begin{eqnarray}
& &B_7^2=(x_B-r^2)M_B^2\;,\;\;\;\; C_7^2=x_KM_B^2
\nonumber\\
& &t_{7a}=\max(A,\sqrt{|B_7^2|},1/b_B,1/b_K)\;,
\nonumber\\
& &t_{7b}=\max(A,C_7,1/b_B,1/b_K)\;.
\end{eqnarray}
The function $F_7$ denotes the product
\begin{eqnarray}
F_7(t)=\alpha_s(t)c_7(t)\exp[-S(x_B,x_K,t,b_B,b_K)]\;.
\end{eqnarray}
The hard functions
\begin{eqnarray}
& &H^{(a)}_{7}(Ab_K,\sqrt{|B_7^2|}b_B,\sqrt{|B_7^2|}b_K)=
\nonumber\\
& &\hspace{1.0cm} K_0(Ab_k)
h(\sqrt{|B_7^2|}b_B,\sqrt{|B_7^2|}b_K)\;,\;\;\;\;B_7^2>0\;,
\nonumber\\
& &\hspace{1.0cm} K_0(Ab_k)
h'(\sqrt{|B_7^2|}b_B,\sqrt{|B_7^2|}b_K)\;,\;\;\;\;B_7^2<0\;,
\end{eqnarray}
with
\begin{eqnarray}
h&=&\theta(b_B-b_K)K_0(\sqrt{|B_7^2|}b_B)
I_0(\sqrt{|B_7^2|}b_K)+(b_B\leftrightarrow b_K)\;,
\nonumber\\
h'&=&i\frac{\pi}{2}\left[\theta(b_B-b_K)
H_0^{(1)}(\sqrt{|B_7^2|}b_B)J_0(\sqrt{|B_7^2|}b_K)+
(b_B\leftrightarrow b_K)\right]\;,
\end{eqnarray}
and 
\begin{eqnarray}
H^{(b)}_{7}(Ab_B,C_7b_B,C_7b_K)=K_0(Ab_B)
h(C_7b_B,C_7b_K)\;,
\end{eqnarray}
come from the two diagrams.

Four diagrams are associated with the operator $O_8$, where the photon is
radiated by each quark in the $B$ and $K^*$ mesons. The corresponding
amplitude is
\begin{eqnarray}
M_{18}&=&-\Gamma^{(0)}\frac{1}{3}\int_0^1dx_Bdx_K\int_0^{1/\Lambda}
b_Bdb_Bb_Kdb_K\phi_B(x_B)\phi_{K^*}(x_K)
\nonumber\\
& &\times \Bigg\{(1-r^2+x_B)(rx_K+x_B)
H^{(a)}_8(Ab_K,B_8b_B,B_8b_K)F_8(t_{8a})
\nonumber\\
& &\hspace{0.5cm}+[(2-3r)x_K-x_B+r(1-x_K)(rx_K-2rx_B+3x_B)]
\nonumber\\
& &\hspace{1.0cm}\times
H^{(b)}_8(Ab_B,C_8b_B,C_8b_K)F_8(t_{8b})
\nonumber\\
& &\hspace{0.5cm}+(1+r)(1-r^2)[(1+r)x_B-rx_K]
\nonumber\\
& &\hspace{1.0cm}\times
H^{(c)}_8(\sqrt{|A'^2|}b_K,D_8b_B,D_8b_K)F_8(t_{8c})
\nonumber\\
& &\hspace{0.5cm}
-[(1-r^2)((1-r^2)(2-x_K)+(1+3r)(2x_K-x_B))
\nonumber\\
& &\hspace{1.0cm}
+2r^2x_K(x_K-x_B)]
\nonumber\\
& &\hspace{1.0cm}\times
H^{(d)}_8(\sqrt{|A'^2|}b_B,E_8b_B,E_8b_K)F_8(t_{8d})\Bigg\}\;
\end{eqnarray}
with
\begin{eqnarray}
& &A^{\prime 2}=(1-r^2)(x_B-x_K)M_B^2\;,\;\;\;\;
B_8^2=(1-r^2+x_B)M_B^2\;,
\nonumber\\
& &C_8^2=(1-x_K)M_B^2\;,\;\;\;\;
D_8^2=(1-r^2)x_BM_B^2\;,\;\;\;\; E_8^2=(1-r^2)x_KM_B^2\;,
\nonumber\\
& &t_{8a}=\max(A,B_8,1/b_B,1/b_K)\;,
\nonumber\\
& &t_{8b}=\max(A,C_8,1/b_B,1/b_K)\;,
\nonumber\\
& &t_{8c}=\max(\sqrt{|A'^2|},D_8,1/b_B,1/b_K)\;,
\nonumber\\
& &t_{8d}=\max(\sqrt{|A'^2|},E_8,1/b_B,1/b_K)\;.
\end{eqnarray}
The funciton $F_8$ denotes the product
\begin{eqnarray}
F_8(t)=\alpha_s(t)c_8(t)\exp[-S(x_B,x_K,t,b_B,b_K)]\;.
\end{eqnarray}
The hard functions
\begin{eqnarray}
& &H^{(a)}_{8}(Ab_K,B_8b_B,B_8b_K)=
H^{(b)}_{7}(Ab_K,B_8b_B,B_8b_K)\;,
\nonumber\\
& &H^{(b)}_8(Ab_B,C_8b_B,C_8b_K)=K_0(Ab_B)h'(C_8b_B,C_8b_K)
\nonumber\\
& &H^{(c)}_8(\sqrt{|A'^2|}b_K,D_8b_B,D_8b_K)=
\nonumber\\
& &\hspace{1.0cm} K_0(\sqrt{|A'^2|}b_K)h(D_8b_B,D_8b_K)\;,
\;\;\;\;A'^2\ge 0\;,
\nonumber\\
& &\hspace{1.0cm} i\frac{\pi}{2}H_0^{(1)}
(\sqrt{|A'^2|}b_K)h(D_8b_B,D_8b_K)\;,\;\;\;\;A'^2< 0\;,
\nonumber\\
& &H^{(d)}_8(\sqrt{|A'^2|}b_B,E_8b_B,E_8b_K)=
\nonumber\\
& &\hspace{1.0cm} K_0(\sqrt{|A'^2|}b_B)h'(E_8b_B,E_8b_K)\;,
\;\;\;\;A'^2\ge 0\;,
\nonumber\\
& &\hspace{1.0cm} i\frac{\pi}{2}H_0^{(1)}
(\sqrt{|A'^2|}b_B)h'(E_8b_B,E_8b_K)\;,\;\;\;\;A'^2< 0\;
\end{eqnarray}
are derived from the four diagrams.

The amplitudes with the structure $i\epsilon_{\mu\rho+-}
\epsilon_\gamma^\mu\epsilon^{*\rho}_{K^*}$ are
\begin{eqnarray}
M_{22}&=&\Gamma^{(0)}\frac{4}{3}\int_0^1\int_0^{1-x}dy\int_0^1dx_Bdx_K
\int_0^{1/\Lambda}bdb\phi_B(x_B)\phi_{K^*}(x_K)
\nonumber\\
& &\times \alpha_s(t_2)c_2(t_2)\exp[-S(x_B,x_K,t_2,b,b)]
\nonumber\\
& &\times \left[\left((1-r)(1-r^2)+2r^2x_K+2x_B\right)y\right.
\nonumber\\
& &\left.-\left(r(1+r)x_K+(3-r)x_B\right)(1-x)\right]
\nonumber\\
& &\times 
\frac{(1-r^2)x_Kx}{xy(1-r^2)x_KM_B^2-M_c^2}H_{2}(Ab,\sqrt{|B_2^2|}b)\;,
\\
M_{27}&=&-M_{17}\;,
\\
M_{28}&=&\Gamma^{(0)}\frac{1}{3}\int_0^1dx_Bdx_K\int_0^{1/\Lambda}
b_Bdb_Bb_Kdb_K\phi_B(x_B)\phi_{K^*}(x_K)
\nonumber\\
& &\times \Bigg\{(1-r^2+x_B)(rx_K+x_B)
H^{(a)}_8(Ab_K,B_8b_B,B_8b_K)F_8(t_{8a})
\nonumber\\
& &\hspace{0.5cm}+[(2-3r)x_K-x_B-r(1-x_K)(rx_K-2rx_B+3x_B)]
\nonumber\\
& &\hspace{1.0cm}\times
H^{(b)}_8(Ab_B,C_8b_B,C_8b_K)F_8(t_{8b})
\nonumber\\
& &\hspace{0.5cm}+(1-r)(1-r^2)[(1+r)x_B-rx_K]
\nonumber\\
& &\hspace{1.0cm}\times
H^{(c)}_8(\sqrt{|A'^2|}b_K,D_8b_B,D_8b_K)F(t_{8c})
\nonumber\\
& &\hspace{0.5cm}-[(1-r^2)((1-r^2)(2+x_K)-(1-3r)x_B)
\nonumber\\
& &\hspace{1.0cm}
-2r^2x_K(x_K-x_B)]
\nonumber\\
& &\hspace{1.0cm}\times
H^{(d)}_8(\sqrt{|A'^2|}b_B,E_8b_B,E_8b_K)F(t_{8d})\Bigg\}\;.
\end{eqnarray}
It is obvious that the above factorization formulas bear the
features of Eq.~(\ref{for}).

The exponentials $\exp(-S)$ appearing in $M_{ij}$
are the complete Sudakov factor, with the exponent
\begin{eqnarray}
S&=&s(x_BP_1^-,b_B)+2\int_{1/b_B}^{t}\frac{d\mu}{\mu}\gamma_q(\alpha_s(\mu))
\nonumber\\
& &+s(x_KP_2^+,b_K)+s((1-x_K)P_2^+,b_K)
+2\int_{1/b_K}^{t}\frac{d\mu}{\mu}\gamma_q(\alpha_s(\mu)).
\end{eqnarray}
The wave functions $\phi_B$ and $\phi_{K^*}$ satisfy the normalization,
\begin{equation}
\int_0^1\phi_i(x)dx=\frac{f_i}{2\sqrt{6}}\;,
\end{equation}
with the decay constant $f_i$, $i=B$ and $K^*$. The wave funciton for the
$K^*$ meson with transverse polarizations has been derived using QCD sum
rules \cite{CZ}, given by
\begin{equation}
\phi_{K^*}=\sqrt{6}f_{K^*}x(1-x)[0.7-(1-2x)^2]\;.
\end{equation}
As to the $B$ meson wave functions, we investigate two models \cite{YS,BW},
\begin{eqnarray}
\phi^{(I)}_B(x)&=&\frac{N_Bx(1-x)^2}{M_B^2+C_B(1-x)}\;,
\label{bw}\\
\phi^{(II)}_B(x)&=&N'_B\sqrt{x(1-x)}
\exp\left[-\frac{1}{2}\left(\frac{xM_B}{\omega}\right)^2\right]\;,
\label{os}
\end{eqnarray}
with the normalization constants $N_B$ and $N'_B$, and the shape parameters
$C_B$ and $\omega$.

\vskip 1.0cm

\centerline{\large \bf 4. Results}
\vskip 0.5cm

In the evaluation of the various form factors and amplitudes, we adopt
$G_F=1.16639\times 10^{-5}$ GeV$^{-2}$, the decay constants $f_B=200$ MeV
and $f_{K^*}=220$ MeV, the CKM matrix elements
$|V^*_{ts}V_{tb}|=0.04$, the masses $M_c=1.5$ GeV, $M_B=5.28$ GeV and
$M_{K^*}=0.892$ GeV, and the ${\bar B}^0$ meson lifetime
$\tau_{B^0}=1.53$ ps \cite{YL}. It turns that no matter what value of the
shape parameter $C_B$ is chosen, the model $\phi^{(I)}_B$ in
Eq.~(\ref{bw}) with a flat profile leads to results much smaller than
the CLEO data of the branching ratio ${\cal B}(B\to K^*\gamma)=
(4.2\pm 0.8 \pm 0.6) \times 10^{-5}$ \cite{Stone}:
the maximal preiction from $\phi^{(I)}_B$, corrsponding to the shape
parameter $C_B=-M_B^2$, is about $1.1\times 10^{-5}$.
The model $\phi^{(II)}_B$ in Eq.~(\ref{os}) with a sharp peak at small $x$
then serves a possible candidate. It is indeed found that as $\omega=0.425$
GeV, a prediction $4.19\times 10^{-5}$ for the branching ratio is reached,
which is equal to the central value of the experimental data. If varying
the shape parameter to $\omega=0.42$ GeV and to $\omega=0.43$ GeV,
we obtain the branching ratios $4.36\times 10^{-5}$ and $4.08\times 10^{-5}$,
respectively. It indicates that the allowed range of $\omega$ is wide due
to the large uncertainties of the data.

The detailed contribution from each amplitude $M_{ij}$ is listed in the
following table in the unit of $10^{-6}$ GeV$^{-2}$:
\[\begin{array}{ccc}\hline
M_{12}/\Gamma_0 & M_{17}/\Gamma_0 & M_{18}/\Gamma_0 \\ 
\hline
-6.20-20.38i & 11.10-198.08i & -1.94-0.29i \\
\hline
M_{22}/\Gamma_0 & M_{27}/\Gamma_0 & M_{28}/\Gamma_0 \\ 
\hline
-4.84-16.17i & -11.10+198.08i & 2.13+0.36i \\
\hline 
\end{array}\]
It is observed that the contributions from the operator $O_2$ are 
comparable with those from the penguin operator $O_7$. The amplitudes
associated with $O_8$ are less important. The other operators give
contributions smaller than those from $O_8$ by an order of magnitude,
which are thus not included here. Furthermore, the imaginary
parts of $M_{1(2)7}$ from the timelike interal quarks play an essential
role for the explanation of the data.

In this work we have determined the $B$ meson wave function,
\begin{equation}
\phi_B(x)=1.81359\sqrt{x(1-x)}
\exp\left[-\frac{1}{2}\left(\frac{xM_B}{0.425\;\;{\rm GeV}}\right)^2
\right]\;,
\label{bwf}
\end{equation}
from the best fit to the experimental data of ${\cal B}(B\to K^*\gamma)$.
We stress that, however, Eq.~(\ref{bwf}) is not conclusive because of
the large allowed range of the shape parameter $\omega$. A more precise
$B$ meson wave function can be obtained by considering a global fit to the
data of various decay modes, including $B\to D^{(*)}\pi(\rho)$ \cite{LM}.
Once the $B$ meson wave fucntion is fixed, we shall employ it
in the evaluation of the nonleptonic charmless decays.

\vskip 0.5cm

This work was supported by the National Science Council of Republic of
China under the Grant No. NSC-88-2112-M006-013.

\end{document}